\documentclass[twocolumn,floats,showpacs,amsmath,amssymb,pre]{revtex4}
\usepackage{graphicx}
\begin{document}

\title{Zero-temperature Glauber dynamics on small-world networks}
\author{Carlos P. Herrero}
\affiliation{Instituto de Ciencia de Materiales de Madrid,
         Consejo Superior de Investigaciones Cient\'{\i}ficas (CSIC),
         Campus de Cantoblanco, 28049 Madrid, Spain }
\date{\today}

\begin{abstract}
The zero-temperature Glauber dynamics of the ferromagnetic Ising
model on small-world networks, rewired from a two-dimensional square
lattice, has been studied by numerical simulations.
For increasing disorder in finite networks, the nonequilibrium dynamics 
becomes faster, so that the ground state is found more likely.
For any finite value of the rewiring probability $p$, the likelihood 
of reaching the ground state goes to zero in the thermodynamic limit,
similarly to random networks.
The spin correlation $\xi(r)$ is found to decrease with distance as 
$\xi(r) \sim \exp(-r/\lambda)$, $\lambda$ being a correlation length
scaling with $p$ as $\lambda \sim p^{-0.73}$. These results are
compared with those obtained earlier for addition-type small world 
networks.
\end{abstract}
%

\maketitle

\section{Introduction}

In recent years, researchers have accumulated evidence that various
kinds of complex systems can be described in terms of networks or graphs,
where nodes play the role of system units and edges represent
interactions between connected pairs of units.
Thus, complex networks have been used to model several types of real-life
systems, and to study various
processes taking place on them \cite{al02,do03a,ne03,ne06,co07}.
In this context, some models of networks have been designed to explain
empirical data in several fields (sociology, economy, biology, technology),
as is the case of the so-called small-world networks, introduced by 
Watts and Strogatz in 1998 \cite{wa98}. 

This kind of small-world networks are well suited to describe properties 
of systems with underlying topological structure ranging from
regular lattices to random graphs \cite{bo98,ca00}, 
by changing a single parameter \cite{wa99}.
They consist of a regular lattice,
in which a fraction $p$ of the bonds between nearest-neighbor nodes
are replaced by new random links, thus creating long-range
``shortcuts'' \cite{wa98,wa99}.  This procedure generates networks
in which one finds at the same time a local neighborhood
(as in regular lattices) and some global properties of random graphs,
as a small mean distance between pairs of sites.
This short global length scale has been found to be relevant for
several statistical physical problems on small-world networks,
such as spread of infections \cite{ku01,mo00a},
signal propagation \cite{wa98,la00,he02b,mo04}, and
information spreading \cite{la01b,pa01,la01a,he03,he07,ca07}. 
                    
Cooperative phenomena in this kind of networks display 
unusual characteristics, associated to their peculiar topology 
\cite{ba00,ne00a,mo00b,sv02,ca06,do08,he08}. 
Thus, a paramagnetic-ferromagnetic phase transition of mean-field
type at finite temperature was found for the Ising model on 
small-world networks derived 
from one-dimensional (1d) \cite{ba00,gi00,lo04}, as well as from
2d and 3d regular lattices \cite{he02a,ha03}.

While the single-spin-flip Glauber dynamics at finite temperatures 
reaches the state given by thermodynamic equilibrium, the situation at 
zero temperature 
is not so clear. In fact, it has been shown that nontrivial phenomena
appear, even for regular lattices \cite{br89,br94,sp01,sp01b}.
In $d=1$ one reaches the ordered ground state, but for higher dimensions
the system may get stuck in a frozen state with domains of opposite 
magnetization \cite{sp01b}. The Glauber dynamics has been also 
investigated on various types of complex networks, in particular
in small-world networks \cite{bo03,da05},
Erd\"os-R\'enyi random graphs \cite{ha02,ca05,uc07},
and scale-free networks \cite{ca05,zh05,ca06b,uc07}. In these cases, 
the system may become trapped in a set of ordered domains, 
without reaching a fully ordered state, even at finite system size.
Analytical calculations, in particular,  have been carried out to study 
several characteristics of this zero-temperature dynamics on regular 
lattices \cite{br89,br94,go05,ne99b} and complex networks
\cite{ca05,ca06b,ha02}. A comparison with numerical simulations 
has shown that mean-field-type descriptions may be inappropriate to 
describe this problem \cite{ca06b}.

Here we consider the zero-temperature Glauber dynamics in small-world
networks rewired from a 2d square lattice. This allows us to study
the evolution of the system from a regular lattice to a random 
network by changing the rewiring probability $p$. 
Earlier works considered networks in which links were added
to the regular lattice, so that a random network was not approached
in the limit of large disorder \cite{bo03,da05}. We will look for
similarities and differences between the spin dynamics on these 
``addition-type'' networks and those studied here.
A question of particular interest is whether for small disorder
($p \ll 1$) the system behaves like a random network, or on the
contrary it keeps some characteristics of the regular lattice.  

The paper is organized as follows.
In Sec.~II we describe the model employed in the paper.
In Sec.\,III we discuss the ordering process, presenting results for
the fraction of ordered and active runs, as well as the ordering time.
In Sec.\,IV we give results for the active links,
in Sec.\,V we present the spin correlation, and 
in Sec.\,VI we discuss the effect of the initial magnetization on
the system evolution.
The paper closes with a summary in Sec.\,VII.

\section{Model}

To generate our networks, we start from a two-dimensional square 
lattice. Contrary to the 1d chain, it is known that in 2d lattices
the zero-temperature Glauber dynamics may get stuck in a frozen state 
without reaching the ground state (a ferromagnetic state, with all
spins parallel), which makes the dynamics nontrivial \cite{sp01,sp01b}.
Small-world networks were built up according to the model
of Watts and Strogatz \cite{wa98,wa99}, i.e.,
we considered in turn each of the bonds in the starting 2d lattice and 
replaced it with a given probability $p$ by a new connection.
In this rewiring process, one end of the selected link is changed
to a new node chosen at random in the whole network.
We impose the conditions: (i) no two nodes can have more than
one bond connecting them, (ii) no node can be connected by a link
to itself, and (iii) each node has at least two connections.  
With this method we obtained networks where 
more than 99.9\% of the sites were connected in a single component.
Moreover, this rewiring procedure keeps constant the total number of links 
in the rewired networks, so that we have an average degree 
$\langle k \rangle = 4$ irrespective of the rewiring probability $p$.
This allows us to study the effect of disorder
upon the properties of the model, without changing the mean
connectivity.
For networks generated in the present way there is a $p$-dependent 
crossover size that separates the large- and small-world regimes, and 
the small-world behavior appears for any finite value of $p$ ($0<p<1$) 
as soon as the network is large enough \cite{ba99c,me00}.
All networks considered here are in the small-world 
regime \cite{he02a,he02b}.

We note that other ways of generating small-world networks from regular
lattices have been proposed \cite{ne99,ne00b}. In particular, instead of 
rewiring each bond with probability $p$, one can add shortcuts between
pairs of sites taken at random, without removing bonds
from the regular lattice. This method turns out to be more convenient
for analytical calculations, but does not keep constant the mean
degree $\langle k \rangle$, which in this case increases with $p$.
The zero-temperature Glauber dynamics of the Ising model has been studied
earlier on these ``addition-type'' small-world networks. Thus,
Boyer and Miramontes \cite{bo03} studied interface motion and pinning 
in the limit $p \ll 1$, whereas Das and Sen \cite{da05} considered densely 
connected small-world networks, generated from one-dimensional rings.

From the 2d square lattice, we generated small-world networks of different
sizes $N = L \times L$. For convenience, we will employ later $L$ instead of
$N$ to indicate the system size in some plots. The largest networks used 
here included $N = 400 \times 400$ nodes. 
We considered periodic boundary conditions for the starting regular
lattice, and then the rewiring process was carried out in a way similar
to that described in Ref. \cite{wa98}.
Each network is characterized by its adjacency matrix $A_{ij}$,
where $A_{ij} = 1$ if nodes $i$ and $j$ are connected by a link,
and $A_{ij} = 0$ otherwise.
For a given network, we consider an Ising model with spin variables
$S_i = \pm 1$ ($i = 1, ..., N$) located on the nodes of the network, 
i.e., we have a Hamiltonian:
\begin{equation}
H = - \sum_{i < j} A_{ij} S_i S_j   \, .
\end{equation}
This means that each edge in the network represents a ferromagnetic 
interaction between spins on the two linked nodes.
The spin configuration evolves in time, and at a given time $t$,
each site $i$ experiences a local field $h_i(t)$ due to the
spins located at its nearest-neighbor nodes:
\begin{equation}
h_i(t) = \sum_j A_{ij} S_j(t)   \, .
\end{equation}
The zero-temperature Glauber dynamics is then defined as follows.
At time step $t+1$ one selects at random a node from a uniform
probability distribution, i.e. all nodes are equally likely to be 
selected, irrespective of the particular properties of each one,
such as its degree $k$.
Note that other ways of choosing the nodes, as a deterministic sweep,
or according to a degree-dependent probability distribution,
introduce an undesired bias in the spin dynamics.
Once a node is selected, the value of its spin is updated according 
to the local field, namely:
\begin{equation}
S_i(t+1) =  \left\{
     \begin{array}{ll}
         +1 & \mbox{ if } h_i(t) > 0  \\
         -1 & \mbox{ if } h_i(t) < 0  \\
         \pm 1 & \mbox{ with probability } \frac12 \mbox{ if } h_i(t) = 0.
     \end{array}
\right.
\label{Sij}
\end{equation}

We will measure time as the number of attempted updates per node,
$N_s = t / N$ (simulation sweeps), so that on average each node tries to 
change its state once per simulation sweep.
Several variables characterizing the considered
model have been calculated and averaged for different values of $p$.
In general, we have considered 1000 simulation runs for each rewiring 
probability $p$, but in some cases we carried out up to 5000 runs to 
improve the precision of our results, in particular to determine the
average number of active links (see below). 
We started the simulations by assigning to each node a spin $+1$ or $-1$
with 50\% probability. In Sect. V, however, we consider the effect of
having a starting finite magnetization on the evolution of the system.
For each run considered, the simulation proceeds while the energy of the
system is changing, and stops when it is constant for a long simulation 
interval. This interval is taken to be equal to $2000 N$ attempted spin 
updates, i.e., 2000 simulation sweeps. This is what we will call in the
following the infinite-time limit, to distinguish it from the simulation
times at which the system energy is still changing.

\section{Ordering process}

\begin{figure}
\vspace{-2.0cm}
\hspace{-0.5cm}
\includegraphics[width= 9cm]{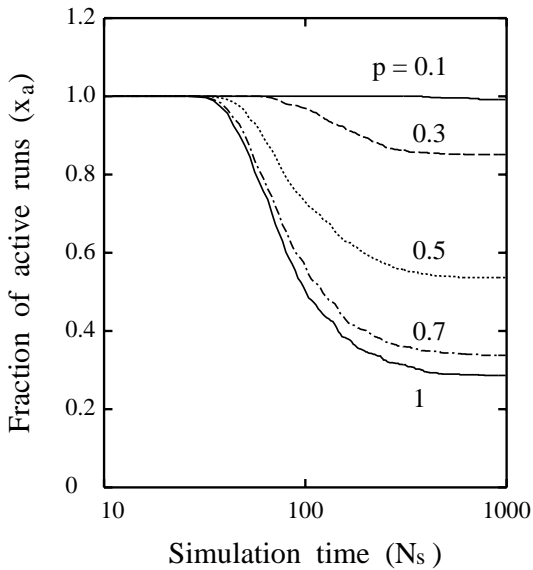}
\vspace{-2.5cm}
 \caption{
Fraction of active spin configurations vs. number of simulation
sweeps, $N_s$, for several values of the rewiring probability $p$.
From top to bottom: $p$ = 0.1, 0.3, 0.5, 0.7, and 1.
Data correspond to simulations on networks with 6400 nodes.
} \label{fig1} \end{figure}

For the zero-temperature Glauber dynamics on
the 2d square lattice, it is known that the system reaches either
a frozen stripe state with probability $\approx 1/3$ or the ordered ground
state with probability $\approx 2/3$ \cite{sp01b}.
For small-world networks we find that the fraction of ordered configurations
depends markedly on the rewiring probability.
We will call active runs those that have not reached the ground state
at a given simulation time. For a given parameter set ($p, N$) we will
denote the fraction of active runs by $x_a$.
In Fig.~1 we display $x_a$ as a function of the number of simulation
sweeps, for several values of $p$. These results were obtained by averaging
in each case over 1000 spin configurations on networks including 6400 nodes.
From these results we observe that the system gets ordered more frequently
as the rewiring probability increases. 
In fact, for $p = 0.1$ we find at large simulation times a fraction of active
spin configurations larger than 0.99, i.e. a fraction of ordered configurations
less than 0.01. 
By comparing with the known result for the regular lattice ($p = 0$), this
means that a small fraction of rewired links is enough to keep the system 
with a large probability in a disordered configuration.
This is in line with the result found in Ref.~\onlinecite{bo03} for
addition-type small-world networks, in the sense that nodes with long-range 
connections strongly affect the motion of interfaces and thus inhibit reaching 
the ground state.

 For larger $p$ the underlying regular lattice is
progressively destroyed, approaching a random network in the limit
$p \to 1$. For random networks it is known that the long-time behavior of
the model is very different from that of the 2d square lattice.
For such networks the probability of the system reaching the ground state
goes to zero in the large-network limit ($N \to \infty$) \cite{ha02}.
In Fig.~1 one also observes that the typical ordering time for
small-world networks decreases
as the rewiring probability is raised. This is general for different
network sizes, as will be discussed below.

For finite networks, it is interesting to check the crossover from
the properties of the regular lattice to those characterizing the 
small-world regime. This crossover will happen at a rewiring
probability $p_c$ that depends on the system size, and decreases as
the system size rises (for $N \to \infty$, $p_c \to 0$).
Thus, for the system size $N = 6400$ considered in Fig.~1, one has a
change in the fraction of active networks at large time, $x_a$,
from $\approx$ 1/3 for the square lattice ($p=0$) to 0.99 for $p = 0.1$.
This means that there is a fast change of $x_a$ in this region, due to
the onset of the small-world regime. We have checked that, in fact, 
the crossover appears for this system size at $p_c \approx 0.02$.

\begin{figure}
\vspace{-2.0cm}
\hspace{-0.5cm}
\includegraphics[width= 9cm]{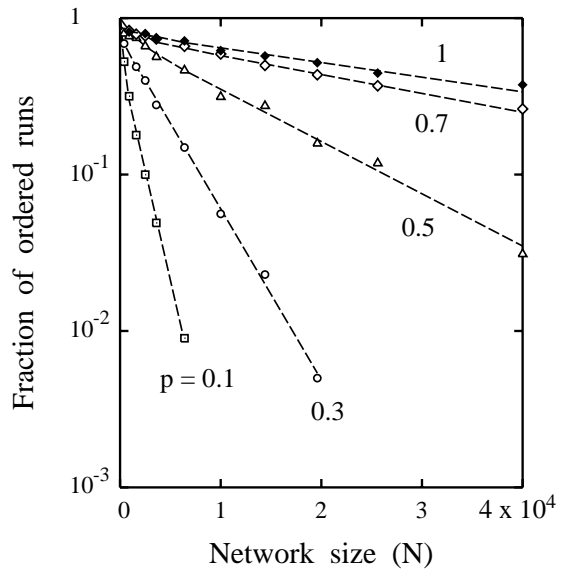}
\vspace{-2.5cm}
\caption{
Fraction of simulation runs reaching the ground state vs. network
size $N$ in the limit of infinite simulation time. Each kind of symbols
indicates a rewiring probability $p$. From top to bottom:
$p$ = 1, 0.7, 0.5, 0.3, and 0.1.  Lines are guides to the eye.
Error bars are on the order of the symbol size.
} \label{fig2} \end{figure}

In Fig.~2 we show the fraction of ordered configurations, $1 - x_a$, 
as a function of the network size $N$, in the limit of large simulation
time. We display in a semilogarithmic plot results for various values 
of the rewiring probability, between $p = 0.1$ and $p = 1$. 
In all cases the probability of a spin configuration reaching the ground 
state decreases exponentially as the network size increases, and the
system remains trapped in spin configurations with part of the nodes
with $S_i = 1$ and the rest with $S_i = -1$.
This behavior is similar to that observed for the Glauber dynamics 
in random networks at $T = 0$.
In fact, Svenson \cite{sv01} has noticed a freezing in a
disordered state for Glauber dynamics on random graphs. This problem
was considered analytically by H\"aggstr\"om, who showed that the
dynamics fails to reach the ordered ground state in random networks 
in the large-size limit $N \to \infty$ \cite{ha02}.
It is important to emphasize the change in $x_a$ from the regular
lattice to small-world networks in the thermodynamic limit. In fact,
$x_a$ changes from $\approx$ 1/3 for the square lattice ($p=0$) to 1 for 
any rewiring probability $p>0$.
The small-world topology thus precludes the system from reaching the 
ordered ferromagnetic state in the large-size limit.
The similarity of our results for rewired small-world networks
with those found for random networks is a difference with addition-type 
networks, since in the latter the connections in the underlying regular 
lattice are preserved for any value of $p$, and one does not recover
the behavior of random networks.

\begin{figure}
\vspace{-2.0cm}
\hspace{-0.5cm}
\includegraphics[width= 9cm]{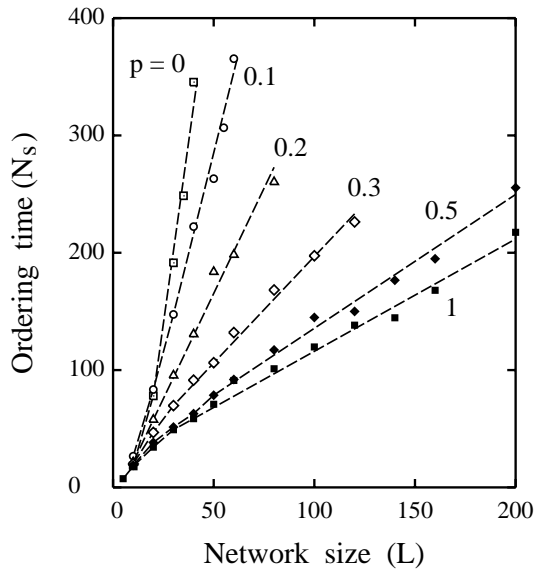}
\vspace{-2.5cm}
\caption{
Mean ordering time as a function of network size for several values of
the rewiring probability $p$. Symbols indicate results derived from
zero-temperature simulations. From left to right: $p$ = 0 , 0.1,
0.2, 0.3, 0.5, and 1.
} \label{fig3} \end{figure}

Another interesting quantity is the mean ordering time of the system, $N_o$.
It is defined as the average simulation time employed by the system to
reach a ground-state configuration, calculated for the runs that
actually become ordered. 
The mean ordering time is displayed in Fig.~3 as a function of the
network size for several values of $p$, between 0 and 1.
For a given rewiring probability, $N_o$ increases linearly with $L$, 
i.e. $N_o \sim \sqrt{N}$.
For a given network size, the system gets ordered faster for larger
$p$. This means that for small $p$ there appear less ordered configurations
(see Fig.~2), and those that appear take longer times to reach the ground
state. 
This is an interesting difference between the behavior of $x_a$ and
$N_o$ for small $p$. For a given network size, $N_o$ evolves in a 
continuous manner from $p = 0$ to $p = 1$, as shown in Fig.~3.
However, the fraction of active runs $x_a$ for $p= 0$ behaves in a way
different from that for $p > 0$, since in the former case $x_a$ converges to
$\approx 1/3$ as the network size increases \cite{sp01b,ol06}, and in
the latter $x_a \to 1$ in the large-size limit (see above).

\section{Active links}

\begin{figure}
\vspace{-2.0cm}
\hspace{-0.5cm}
\includegraphics[width= 9cm]{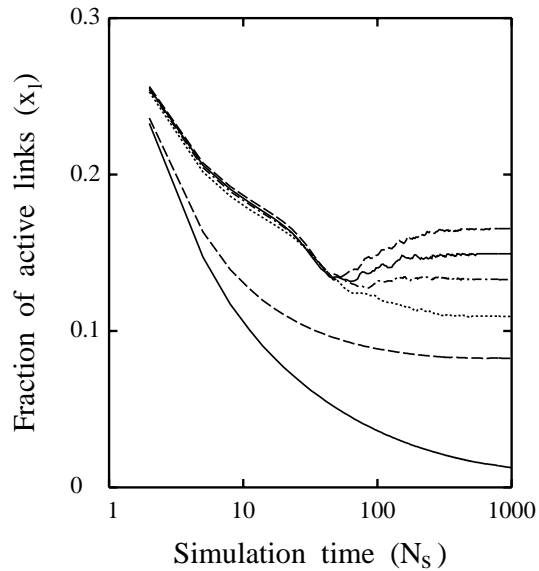}
\vspace{-2.5cm}
\caption{
Fraction of active links in active runs as a function of
the number of simulation sweeps. Lines correspond to various values
of the rewiring probability $p$. From top to bottom: $p$ = 1, 0.7,
0.6, 0.5, 0.1, and 0.
These data correspond to simulations on networks with 6400 nodes.
} \label{fig4} \end{figure}

The failure of the system to reach an ordered state for system size 
$N \to \infty$
could be due to the presence of small clusters (communities) of nodes
tightly connected with each other, but loosely linked to the rest of
the network. Such clusters could become ordered independently of the
rest of the system, thus giving a disordered state, and hindering the
reach of the ground state. 
This is the mechanism discussed in Ref.~\onlinecite{ca05} in connection
with the Glauber dynamics on random networks.
To address the validity of this assumption, it is useful to consider 
the active links in a network, defined as those connecting nodes 
with opposite values of $S_i$.
In the following we will call $x_l$ the fraction of active links, 
averaged over active runs.
The evolution of $x_l$ along the simulation runs is presented in
Fig.~4 for various values of the rewiring probability and networks
including 6400 nodes.
We note first that, for a given simulation time $N_s$, the fraction of 
active links increases as $p$ is raised. For $p \le 0.5$, $x_l$ decreases
as time proceeds and eventually converges to a finite value. 
For $p > 0.5$ we find a decrease in $x_l$ at short times, reaching
a minimum for $N_s \sim 100$, and rising latter to saturate to
a finite value at large times.
In principle, for the ordering process advancing as time proceeds,
the number of actives links in active configurations decreases, as
found for $p < 0.5$.
However, a fast decrease in the fraction of active runs can cause
a rise in $x_l$, since the latter is calculated over runs still
active, and those getting ordered are expected to have few
active links. In fact, the minimum in $x_l$ observed for
$p > 0.5$ occurs at the same simulation times as the fast decrease
in $x_a$ shown in Fig.~1. 
This minimum in $x_l$ is reminiscent of that found by Castellano
{\em et al.} \cite{ca05} for random networks with average degree
$\langle k \rangle$ = 7.

\begin{figure}
\vspace{-2.0cm}
\hspace{-0.5cm}
\includegraphics[width= 9cm]{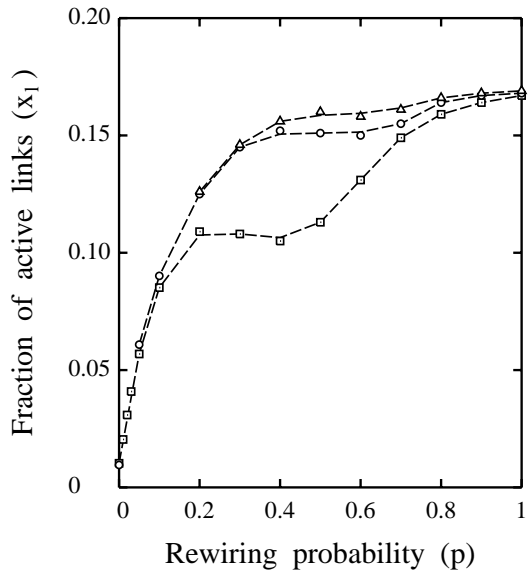}
\vspace{-2.5cm}
\caption{
Fraction of active links in active runs vs rewiring
probability $p$, in the limit of infinite simulation time.
Symbols represent results of simulations for networks with different
sizes: squares, $L$ = 100; circles, $L$ = 300; triangles, $L$ = 400.
Error bars are on the order of the symbol size.
Lines are guides to the eye.
} \label{fig5} \end{figure}

The long-time value of $x_l$ is presented in Fig.~5 as a function
of the rewiring probability $p$ for three different system sizes.
Symbols indicate results of our simulations: squares for $L$ = 100,
circles for $L$ = 300, and triangles for $L$ = 400. 
For $L$ = 100 we find that $x_l$ increases with $p$, but has a 
plateau at intermediate values of the rewiring probability.
This plateau, however, seems to be a finite-size effect, since
it tends to disappear as the network size is increased.
In fact, for $L$ = 400, it is almost inappreciable; 
one has a fast rise in the fraction of active links $x_l$
between $p$ = 0 and $p \approx 0.3$, and the rise is much slower for 
$p > 0.4$. Finally, in the large-disorder limit ($p = 1$) $x_l$
converges to a value of about 0.17, irrespective of the system size $L$.
Interestingly, a clear finite-size effect appears for intermediate
values of $p$, but is very small in both limits $p \to 0$
(regular lattice) and $p \to 1$ (random networks).  
Close to $p = 0$ we find a linear increase in $x_l$, with a slope
$d x_l/d p = 1.02 \pm 0.02$. 

We now go back to the question posed above on the
presence of small clusters of nodes that could get ordered independently
of the rest of the network. If this was the reason for avoiding the full
ordered state, one would expect to find a small fraction of active links,
which should decrease for rising network size. This is not the case, as 
shown by our results displayed in Fig.~5. In line with this, we have also 
observed in our simulations that the long-time magnetization goes to zero 
when increasing the system size, as expected for the presence of different 
large ordered regions in the network. This conclusion agrees with that
given in Ref. \onlinecite{ca05} for random networks, but here it
extends to small-world networks in the whole range from $p = 0$ to 1. 

In connection with this, it is important the question whether the disordered
(metastable) state reached after long simulation time is a frozen state or not.
In our simulations we find that the disordered state is active, in the sense
that some spins continue flipping forever, without changing the energy
of the system. Both spin states ($+1$ and $-1$) coexist, with several
connected domains of identical spins. The continuous flipping happens at
the borders of these domains.
This behavior is similar to that observed for regular
lattices for $d > 2$ and for random networks \cite{ca05}. 
This means that a small fraction of rewired links ($p \ll 1$) is enough to
recover the behavior of random networks, instead of the freezing
observed for the Glauber dynamics on the 2d square lattice \cite{sp01b}.

\section{Spin correlation}

\begin{figure}
\vspace{-2.0cm}
\hspace{-0.5cm}
\includegraphics[width= 9cm]{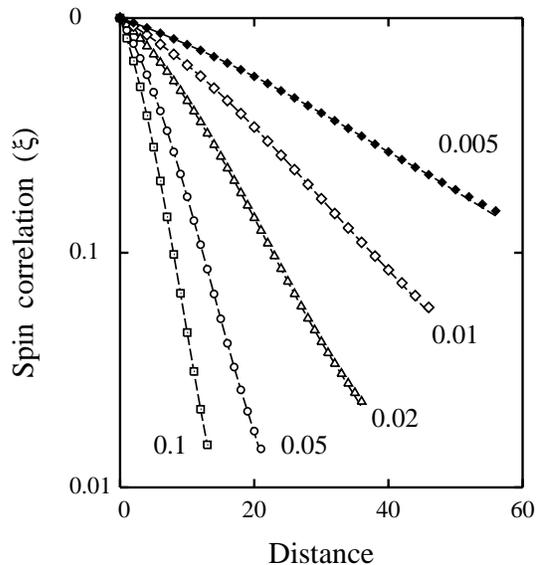}
\vspace{-2.5cm}
\caption{
Spin correlation $\xi(r)$ vs distance $r$ for active runs
on small-world networks with $9\times 10^4$ nodes,
in the limit of large simulation time $N_s \to \infty$.
Error bars are in the order of the symbol size.
} \label{fig6} \end{figure}

From the results presented above, it is clear that the fraction of ordered
configurations on small-world networks goes to zero in the large-size limit, 
and thus the ground state is not reached in the thermodynamic limit.
A direct way of displaying the lack of long-range ferromagnetic ordering 
is by looking at the spin correlation vs distance 
for various values of the rewiring probability $p$.  We define $\xi$ as
\begin{equation}
 \xi(r) =  \langle S_i S_j \rangle_r  \; , 
\label{xi}
\end{equation}
where the subscript $r$ indicates that the average is taken for the 
ensemble of pairs $(i, j)$ of sites at distance $r$.
Note that $r = d/d_0$ refers here to the dimensionless distance between 
sites in the starting regular lattice, not to the actual topological distance 
or minimum number of links between nodes in the rewired networks
($d_0$ is the distance between nearest neighbors).
The correlation $\xi(r)$ in the long-time limit is shown in Fig.~6 for 
several values of the rewiring probability $p$.
After a short transient for small $r$, $\xi(r)$ displays a logarithmic
decrease for increasing distance. At distances longer than those presented
in Fig.~6, $\xi(r)$ saturates to a constant value, which is a finite-size
effect and should disappear in the limit $N \to \infty$. 
Note that the results displayed in Fig.~6 correspond to values of the
rewiring probability $p \ll 1$. For larger $p$, the decrease in $\xi(r)$
as a function of distance is very fast.

\begin{figure}
\vspace{-2.0cm}
\hspace{-0.5cm}
\includegraphics[width= 9cm]{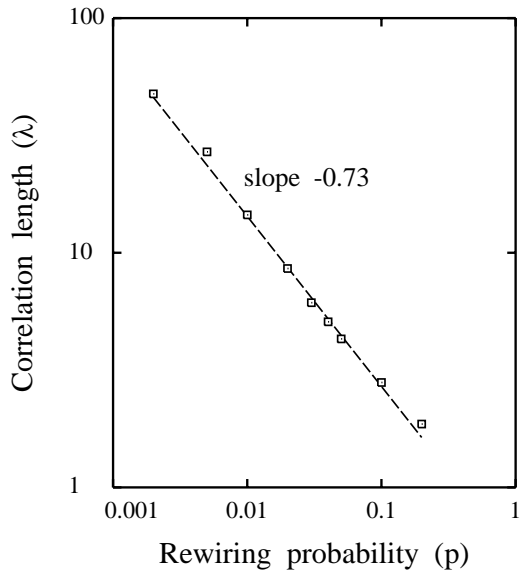}
\vspace{-2.5cm}
\caption{
Correlation length $\lambda$ vs rewiring probability in a logarithmic
plot. Data points were obtained from the decay of the spin correlation
$\xi(r)$ with distance, for networks including $9\times 10^4$ nodes,
in the limit of large simulation time $N_s \to \infty$.
Error bars are on the order of the symbol size.
} \label{fig7} \end{figure}

According to the results shown in Fig.~6, we find a region where the 
spin correlation scales as $\xi(r) \sim \exp(-r/\lambda)$, with a correlation 
length $\lambda$ dependent on the rewiring probability $p$.
This dependence of $\lambda$ on $p$ is shown in Fig.~7 in a logarithmic
plot, where one observes that the correlation length follows over two
decades a power law  $\lambda \sim p^{-a}$, with an exponent 
$a = 0.73 \pm 0.02$.

Boyer and Miramontes \cite{bo03} interpreted the presence of different
spin domains on their addition-type small-world networks in terms of the
``influent'' nodes. These are nodes that have long-range connections,
and strongly affect the motion of domain interfaces.
Moreover, they argued that the exponent $a$ should be 2/3, on the basis
of the random distribution of influent nodes over the square lattice.
In fact, they found a value $a = 0.64$ from their numerical simulations,
close to 2/3.
The exponent yielded by our simulations on rewired networks is somewhat
higher.
This can be understood by taking into account that in our case the
influent nodes play a role similar to that played in addition-type small-world 
networks, but also connections in the underlying lattice are progressively
destroyed as the rewiring probability increases. This affects the correlation 
between sites in the lattice, which is effectively reduced and the exponent
$a$ increases with respect to that expected for the full square lattice
($\lambda$ decreases faster for rising $p$).

\section{Influence of initial conditions}

The results shown above were obtained for initial spin configurations
with zero average magnetization (spins with 50\% probability for $+1$ and
$-1$).  One can also consider starting spin configurations with
different probabilities for spins $+1$ and $-1$, and study the evolution 
of the system under the Glauber dynamics at $T = 0$.  
A study of the effect of initial conditions on this type of dynamics
in various kinds of complex networks has been carried out earlier by
Uchida and Shirayama \cite{uc07}. These authors emphasized that
both the initial conditions and the network structure are equally
relevant to determine the evolution of the system.
This evolution may be nontrivial, mainly in the presence
of nodes with large degree, as happens in scale-free networks. 

\begin{figure}
\vspace{-2.0cm}
\hspace{-0.5cm}
\includegraphics[width= 9cm]{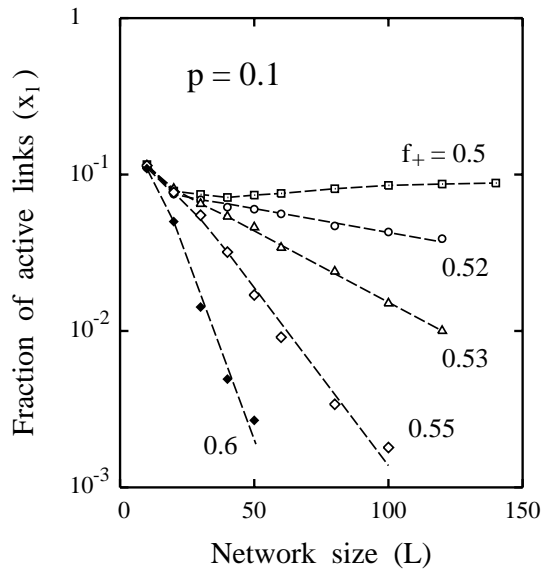}
\vspace{-2.5cm}
\caption{
Fraction of active links vs network size for several values of the
initial magnetization.
Symbols indicate results derived from simulations for a rewiring
probability $p = 0.1$, in the limit of infinite simulation time.
From top to bottom: $f_+$ = 0.5, 0.52, 0.53, 0.55, and 0.6.
For $f_+ > 0.6$, $x_l$ decreases very fast for increasing network
size (not shown).
} \label{fig8} \end{figure}

We now consider initial configurations on our small-world networks,
where the probability $f_+$ of $S_i = +1$ is higher than 0.5, and study 
their evolution with time.
In Fig.~8 we present the fraction of active links as a function of
the network size for several values of $f_+$ between 0.5 and 0.6.
These results were obtained for a rewiring probability $p = 0.1$
and in the long-time limit $N_s \to \infty$.
For $f_+ = 0.5$ (random initial configuration), the fraction $x_l$
converges to a finite value as the network size increases, as 
shown above. For $f_+ > 0.5$, however, $x_l$ decreases exponentially
as $N$ is raised, indicating that the system evolves to a single
ordered domain, eventually converging to the ground state.

\begin{figure}
\vspace{-2.0cm}
\hspace{-0.5cm}
\includegraphics[width= 9cm]{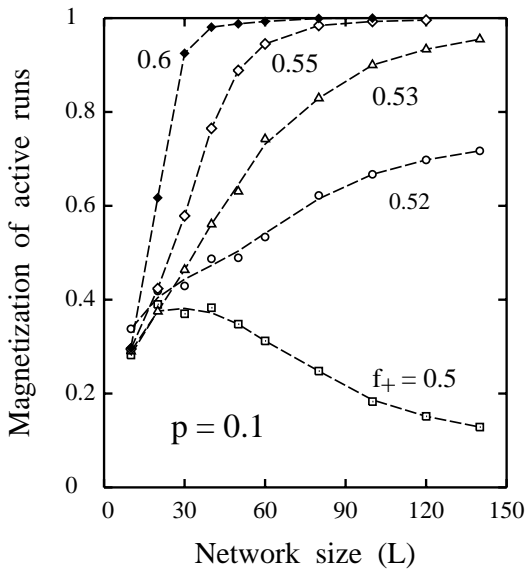}
\vspace{-2.5cm}
\caption{
Average magnetization $\overline{M}$ of spin configurations in
active runs as a function of network size, for various values of
the initial magnetization.
Symbols show results derived from simulations on networks with
a rewiring probability $p = 0.1$, in the limit of infinite
simulation time.
From top to bottom: $f_+$ = 0.6, 0.55, 0.53, 0.52, and 0.5.
For $f_+ > 0.6$, $\overline{M}$ converges fast to 1 for increasing
network size (not shown).
Dashed lines are guides to the eye.
} \label{fig9} \end{figure}

To check this point we have calculated the average magnetization
$\overline{M}$ over active networks as a function of the system size. 
We define $\overline{M}$
as $\overline{M} = \langle | M | \rangle_{ac}$, with $M = \sum_i S_i$.
In Fig.~9 we show $\overline{M}$ for the same rewiring
probability and $f_+$ values presented in Fig.~8. Again the results 
correspond to the long-time limit. 
For random initial conditions the average magnetization decreases and
goes to zero for $N \to \infty$, as indicated above 
(for $L = 300$, we find $\overline{M}$ = 0.057, not shown in the figure).
On the contrary, for $f_+ > 0.5$, the average magnetization increases
as the network size is raised, converging to the value expected for the
fully ordered system ($\overline{M} = 1$).
This is consistent with the decrease observed in the fraction of active
bonds $x_l$ for rising $N$, shown in Fig.~8.
Moreover, this behavior is reminiscent of that observed for random networks 
in Ref. \onlinecite{uc07}, in the sense that a small finite value of the
starting magnetization is enough to drive the system to an ordered state.
Such an evolution can be different for other kinds of complex networks
\cite{uc07}.

Uchida and Shirayama \cite{uc07} found that, depending on the type of 
network, the final dynamics may reach a metastable state involving two 
coexistent spin states with several connected domains of identical
spins, the marginal vertices of which flip continuously.
This is in fact the behavior found for our small-world networks and
presented above for random initial conditions ($f_+ = 0.5$) in section IV.
For $f_+ > 0.5$, however, we have found that the system converges to
an ordered state in the thermodynamic limit $N \to \infty$ 
($\overline{M} \to 1$), but has a finite probability of being 
disordered for any finite size. 
Both Figs.~8 and 9 indicate that, for a given value
$f_+ > 0.5$, the appearance of a single spin domain is favored as 
the system size increases.  However, the convergence to the
ordered state for increasing system size is fast for $p > 0.6$, 
but becomes very slow as one approaches the random initial conditions,
and finally the system remains in a disordered state for $f_+ = 0.5$
and $N \to \infty$ (see above). 
Note that for $f_+ > 0.5$ the behavior of the system is opposite to that 
shown above for $f_+ = 0.5$, in the sense that for random initial conditions
the system reaches the ground state less frequently for increasing
size $N$ (see also the increase in $x_l$ shown in Fig.~8 for $f_+ = 0.5$
and decrease in $\overline{M}$ in Fig.~9, as $N$ rises).

\section{Summary}

We have studied numerically the Glauber dynamics of the ferromagnetic
Ising model on small-world networks, generated by
rewiring links in a two-dimensional square lattice. 
We have found that the behavior of the model departs from that known
for the regular lattice, even for a small fraction of rewired links.
In fact, for any rewiring probability $p$, the fraction of ordered runs
for large time ($N_s \to \infty$) and large system size ($N \to \infty$)
goes to zero, contrary to the regular lattice, where this fraction
converges to a finite value ($\approx 2/3$).
For finite networks, the system gets ordered more frequently as the
rewiring probability increases and one approaches a random network.

The spin correlation on the underlying lattice is found to decrease as
$\xi(r) \sim \exp(-r/\lambda)$, with a correlation length that depends
on the rewiring probability as $\lambda \sim p^{-0.73}$.
The exponent giving the dependence of $\lambda$ on $p$ is close to but 
different from that derived earlier for addition-type small-world networks.

We have analyzed the influence of the initial conditions on the behavior
of the system. We found that for any small deviation from random
initial conditions, the system evolves to the ordered state in the limit
of large system size and simulation time, contrary to the behavior obtained
from random initial conditions.

The apparently simple Glauber dynamics at $T$ = 0 displays a
nontrivial behavior, that depends upon the type of network on
which it is defined. This is true even for regular lattices
with $d > 1$.
For small-world networks rewired from 2d lattices, we find a
behavior of the Glauber dynamics similar to that corresponding
to random networks. This refers in particular to the nature of
the disordered state obtained at long times, which is
characterized by the presence of ordered regions, with spins at
the interfaces between domains flipping forever without
changing the energy. This behavior is different from the frozen
state reached for 2d regular lattices, but is reminiscent of that
for lattices with dimensionality larger than 2.
An important result is that for our small-world networks, a
small fraction of disorder (in fact any rewiring probability
$p > 0$) is enough to recover the behavior known for random
networks.

For small-world networks generated from lattices with a dimensionality
different from 2, we expect a behavior qualitatively similar to that
presented here for $d$ = 2. However, the actual details of the
nonequilibrium ordering process on these networks may depend on $d$.
Also, the behavior of the model for generalized random networks may
display some characteristics different than Erd\"os-R\'enyi networks.
These points require further investigation, and remain as a challenge
for future research.

\begin{acknowledgments}
This work was supported by Ministerio de Ciencia e Innovaci\'on 
(Spain) under Contract No. FIS2006-12117-C04-03.  \\
\end{acknowledgments}

\end{document}